\documentclass[a4paper, 12pt]{article}
\usepackage{amsmath}
\usepackage{amssymb}
\usepackage[utf8]{inputenc}
\usepackage{graphicx}
\usepackage{subfigure}
\def \ni {\noindent}
\def \vs  {\vskip5mm}
\def \bea {\begin{eqnarray}}  
\def \eea {\end{eqnarray}} 
\def \mea {\nonumber\\}
\def \half {{\textstyle \frac{1}{2}}}
\def \third {{\textstyle \frac{1}{3}}}

\begin{document}    
\begin{titlepage}

\title{Mathematics underfoot: \\ The formulas that came to W\"urzburg from New Haven}

\author{ 
A.J. Bracken\footnote{{\em Email:} a.bracken@uq.edu.au}
\\School of Mathematics and Physics\\    
The University of Queensland\\Brisbane 4072, Australia}

\date{}     
\maketitle     
\vs\ni

{\Huge {\em Two formulas are set in floor-tiles in the foyer of the W\"urzburg building that houses the laboratory in which Wilhelm R\"ontgen discovered X-rays in 1895.  But what 
do they mean, and what have they got to do with R\"ontgen or his work?  The answers
involve two distinguished professors and their PhD students, working on opposite sides of the Atlantic
in the $19$th Century.}}

\end{titlepage}    

\setcounter{page}{2}

\section{Playing detective}
On a trip to  the German city of W\"urzburg in 2016 I visited the memorial honouring Wilhelm R\"ontgen, who discovered X-rays there in 1895  and who was awarded
the first Nobel Prize in Physics  in 1901 as a result.\footnote{The story of 
this extraordinary and somewhat accidental discovery 
is well-documented, if not so well-known \cite{roentgen}.  It contrasts sharply with the story behind most recent prize-winning discoveries,
which are typically  the culmination of extended periods of concentrated, well-funded research, aimed at a final goal. }

At the time of his discovery R\"ontgen was 
Professor and Head of the Institute of Physics at Julius-Maximilians-Universit\"at (JMU) W\"urzburg. 
His laboratory is preserved, complete with much of his equipment, in a building that is open to the public and that is
now part of
 the University of Applied Sciences W\"urzburg-Schweinfurt. 
 
   \begin{figure}[htp]
\includegraphics[width=\linewidth]{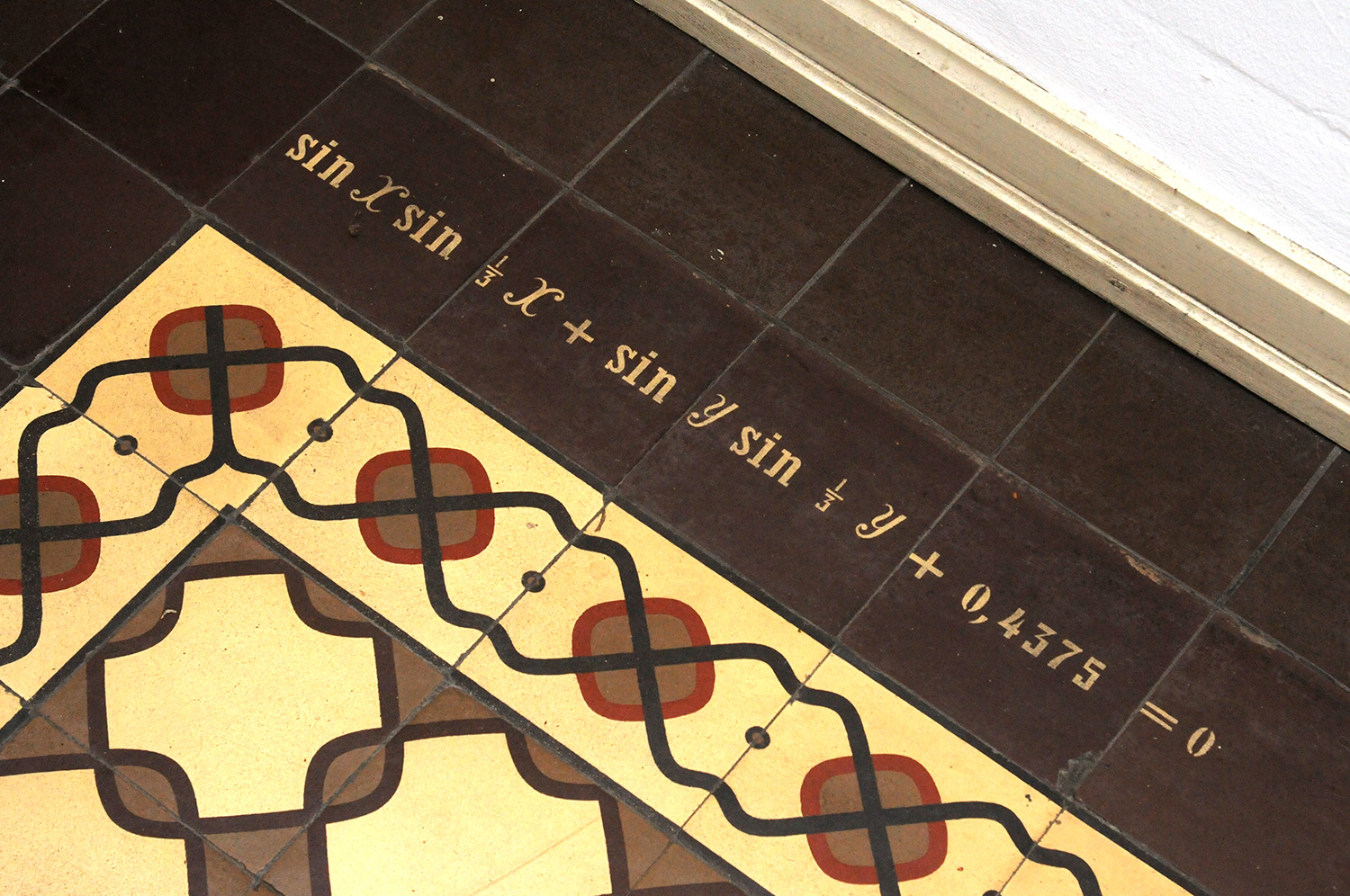}
\caption{Part of the foyer floor, showing the first formula.   Photo courtesy of  G. Bartsch.  }
\end{figure}

 In the
foyer, two trigonometrical formulas are set in the tiled floor.  No explanation is provided for these 
formulas,  and no hint given as to their connection with R\"ontgen's work. 
The two formulas are

\bea
&\,& \quad \sin(x)\,\sin(x/3)+\sin(y)\,\sin(y/3)+0.4375=0\,,
\mea\mea
&\, & 
\quad \sin(y)\,\sin(y/2)=1.3685\, \sin(x)\,\sin(x/3)\,.
\label{formulas1}
\eea

Following my visit, I contacted several colleagues in the 
international  physics and mathematics communities, but none could throw any light on the 
meaning of the formulas.  

\begin{figure}[htp]
\includegraphics[width=\linewidth]{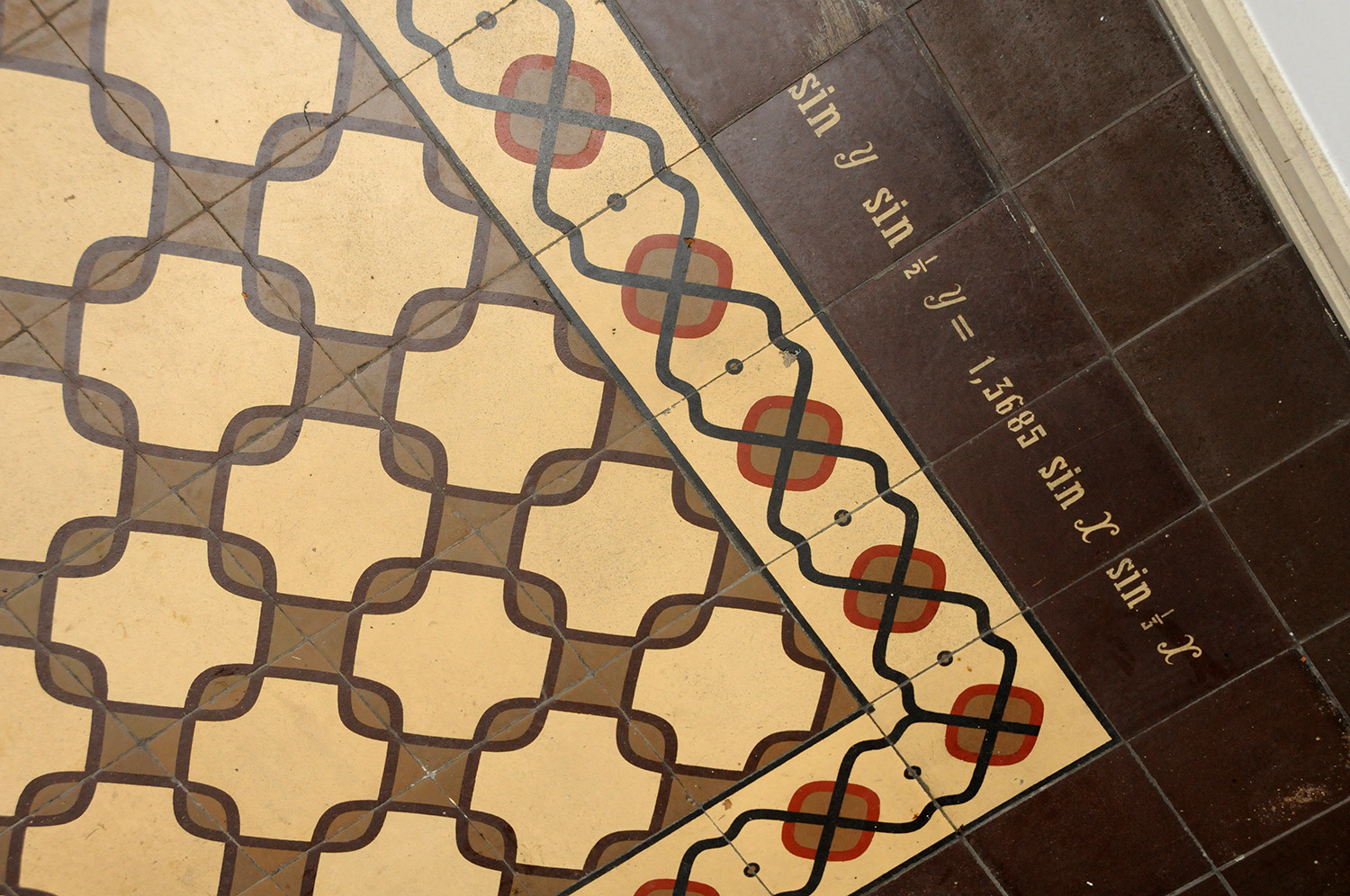}
\caption{Part of the foyer  floor, showing the second formula.  Photo courtesy of G. Bartsch.}
\end{figure}

It occurred to me and  a 
Queensland colleague Dr. Vincent Hart, to treat the formulas as coupled equations in two unknowns,
but this led to pairs of numbers of no apparent  importance.  Frustrated by my continuing ignorance, 
I contacted Mr. Gunnar Bartsch in the Press and Public Relations Section of JMU W\"urzburg, 
but he could not tell me more, though he did supply me with clear pictures of the formulas in the foyer, as in Figs. 1 and 2.  

I must add that when Dr. Peter Jarvis of the University of Tasmania saw these pictures, he 
suggested that the formulas  might be connected 
to the pattern of tiles on the floor next to them, also shown in Figs. 1 and 2.  In view of  what  was revealed later, 
I am somewhat  ashamed to
admit that had not occurred to me;  I had it fixed in my mind that the formulas must have something to do with
R\"ontgen and his work.

The first breakthrough came when
I had a letter  published \cite{bracken1}  seeking
further information from the international physics community about the formulas.  
 This aroused the interest of
Emer. Prof. Steve Webb of the Royal Cancer Hospital, London. 
He asked a German colleague, Prof. Wolfgang Schlegel
of the German Cancer Research Centre in Heidelberg
if he could throw any light on the mystery, and  
Schlegel in turn passed the question to Prof.  Robert Grebner,  President
of the University of Applied Sciences W\"urzburg-Schweinfurt, the present owners of the building.  
As a result Schlegel \cite{schlegel} and in turn Webb and I, were
provided  with a copy of an article  written by journalist Ernst N\"oth about the formulas, published  in the local 
W\"urzburg newspaper {\em Main--Post} back in 1971 \cite{main_post_1971}.  

From this article we learned that the tiled floor
was designed around 1876, 
at the time the building that would eventually house R\"ontgen's laboratory was being planned by 
Friedrich Kohlrausch,  who was R\"ontgen's
predecessor as Professor in the Institute of Physics at JMU W\"urzburg. The tiling was completed together 
with the rest of the building by 1879, and so has nothing
at all to do with R\"ontgen or X-rays.  The two formulas refer to, and in fact define, the patterns on the tiles next to 
them on the floor, as the article makes clear with an accompanying 
computer-generated picture of the curves defined by the first formula.

The article quotes from a hand-written account  by Kohlrausch of the planning of the building 
and its construction, in which he
says that his Research Assistant Vincenc Strouhal got the idea for the floor pattern from 
``an American paper by Newton and Philipps (sic)", and that the tiles
were produced by the firm of ``Villeroy and Koch (sic)"  under the supervision of an engineer named Hrn. Ubach. 

Now I knew the meaning of the formulas and something of their history, 
but was this as much as could be uncovered about them? 
After generating pictures from the two formulas \eqref{formulas1} using {\em MATLAB} \cite{matlab}, 
shown here in Figs. 3 and 4 and closely 
matching the inner and outer
patterns seen on the tiles in Fig. 1 and 2,  I was able to show that 
the numbers $0.4375$ and $1.3685$ appearing there are approximations to exact critical
values $7/16$ and $64\sqrt{3}/81$ at which each of the two sets of curves becomes most closely connected, 
as discussed in Appendices A and B below.  But despite searching
 on the Internet, I could not locate any paper by ``Newton and Philipps (or Phillips or Philips)".

Then came the second major breakthrough.  Gunnar Bartsch published an article 
describing the  story to this point \cite{bartsch}, reproduced in {\em Main-Post} \cite{main_post2}, 
 including a note that  
Anja Schl\"omerkemper, Professor of Mathematics at JMU
W\"urzburg, 
had also confirmed using the computer
that the formulas \eqref{formulas1} do indeed define the inner and outer parts of the pattern on the tiles.  More important
for what followed, 
Bartsch included the remark by Kohlrausch about an ``American paper by Newton and Philipps".
Dr. Matthias Reichling of the Computing Center of JMU W\"urzburg am Hubland saw the article and,
more skilful at searching the Internet than I, located the paper online \cite{newton_phillips} 
and provided me with details.

From this point I was able to discover much of the background about Newton and Kohlrausch and their respective PhD
students, Phillips and Strouhal, and to fill in the story of the tiled floor 
in W\"urzburg as told below, using some judicious guesswork as to how the numbers in the 
formulas \eqref{formulas1} were chosen, and how Strouhal must have designed the  tiles, 
re-working  hand-drawn figures in the paper by Newton and Phillips \cite{newton_phillips}.

Join me now in a trip back in time $\dots$.

\begin{figure}[htp]
\begin{center}
\includegraphics[width=\linewidth, height=4in, trim=0.5in 3.0in 0.1in 3.5in, clip, scale=0.5]{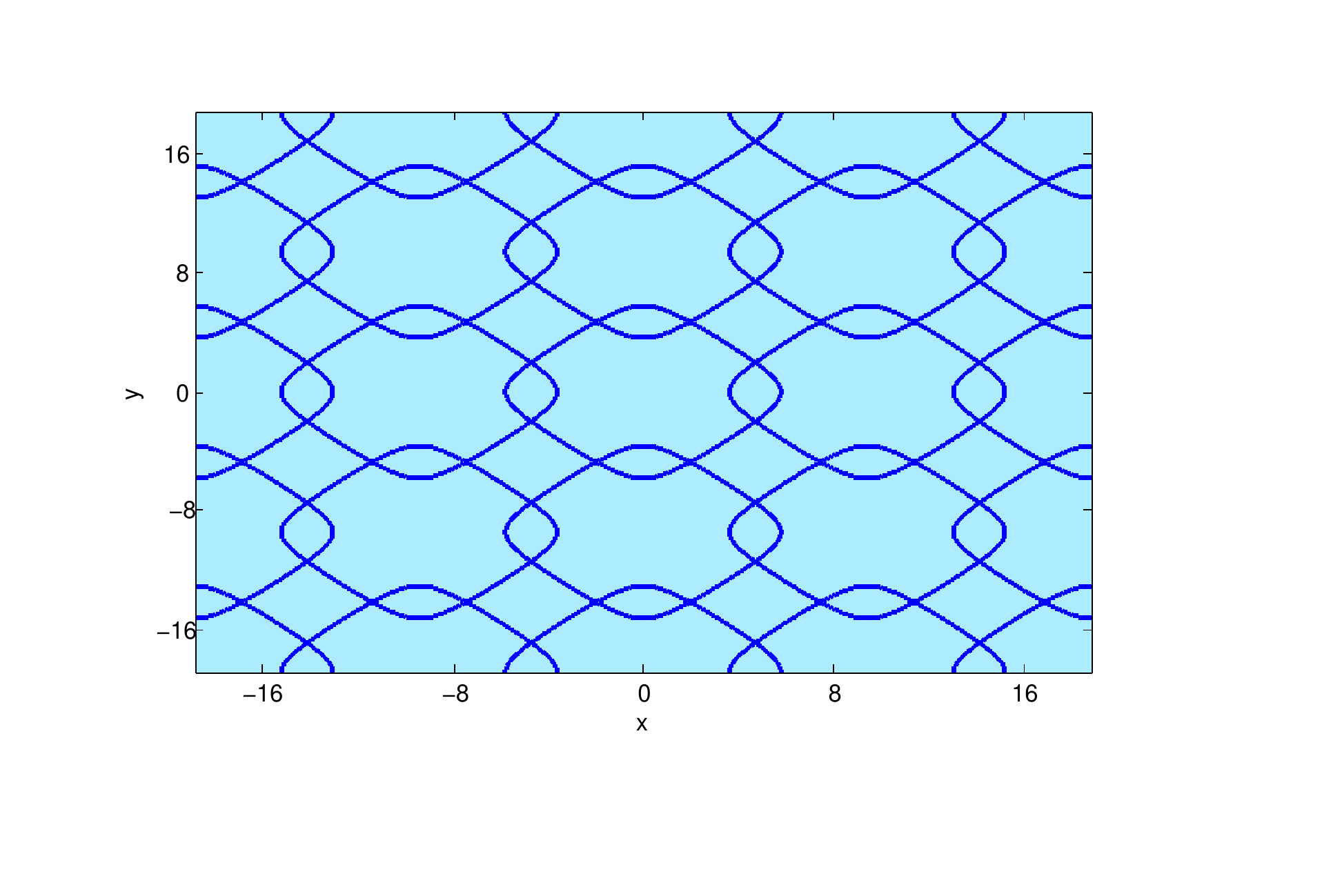}
\end{center}
\setlength{\abovecaptionskip}{-11pt}
\caption{Computer-generated curves defined by the first of formulas \eqref{formulas1},  over four periods in $x$ and $y$, 
from $-6\pi$ to $6\pi$.}
\end{figure}

\section{How it 
may have 
happened}

It is 1875 in New Haven, Connecticut, where  Hubert Anson Newton at age 45 has already been Professor and Head of 
Mathematics at 
Yale University for 20 years, during which he has supervised a number of successful PhD students, including J. Willard Gibbs,
who is now  Professor of Mathematical Physics at Yale.
Newton himself has done important work in pure mathematics and 
actuarial methods, but he also manages the Observatory at Yale,  and has 
established an international reputation for his work on the paths of meteors
and comets \cite{phillips,gibbs}.  He is a Charter Member of the National Academy of Sciences of the United States and is a 
Vice-President, later to be President, of the American Association for the Advancement of Science (AAAS).   
In 1892 he will be elected
a Foreign Member of the Royal Society of London. 

\begin{figure}[htp]
\begin{center}
\includegraphics[width=\linewidth, height=4in, trim=0.5in 3.0in 0.1in 3.5in, clip, scale=0.5]{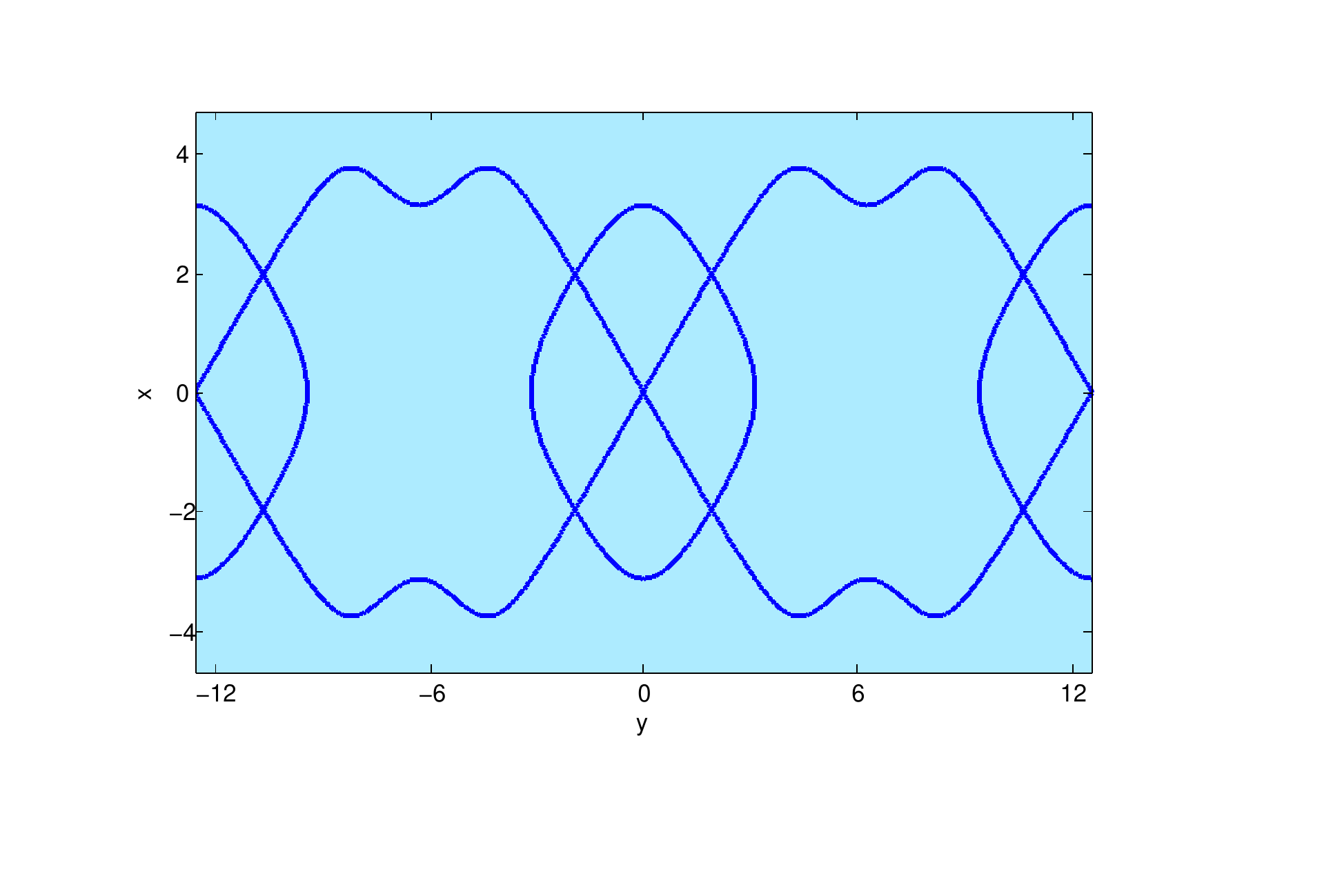}
\end{center}
\setlength{\abovecaptionskip}{-11pt}
\caption{Computer-generated curves defined by the second of formulas \eqref{formulas1}, over one period in $x$,
from $-3\pi/2$ to $3\pi/2$, and two periods in $y$, 
from $-4\pi$ to $4\pi$.}
\end{figure}

By 1875 Newton has 
taken under his wing 31 year old Andrew Wheeler Phillips, 
a gifted student with an unusual mathematical
background.  Until accepted into Newton's Graduate School class without the usual prerequisite of a Bachelor's degree, 
he has been 
largely self-taught \cite{wright}.  Phillips
will complete his PhD thesis 
``On three-bar motion"\footnote{The
three-bar
problem concerns  
the path traced by a  point P fixed relative to the middle bar BC of three straight bars 
AB, BC and CD as they move in a plane.  The bars are  
joined but freely pivoted at B and C,
 and their ends A and D are freely pivoted at fixed points in the plane.}
 in 1877, building on recent work by Roberts \cite{roberts} and Cayley \cite{cayley}, and
 will  later go on  himself to 
become Professor of Mathematics and Dean of the Graduate School at Yale.   During his career, 
he will co-author a variety of influential 
mathematical texts, notably one on 
the graphical interpretation of equations
\cite{phillips5}.

Motivated perhaps  by Newton's work on comets, or by Phillips' thesis topic,  the two co-author a paper in 1875  
in the {\em  Transactions of the Connecticut Academy of Arts and Sciences} (CAAS)
 on the curves in the Cartesian plane defined by certain transcendental equations \cite{newton_phillips}. 
After noting that ``attempts to classify the numerous varieties of transcendental curves have been rare", and that it is not easy from the form of such a curve ``to state  an equation that can represent it",  they remark that ``the simpler inverse problem  
of describing the curve from the equation is naturally the first to be undertaken"  in the hope that ``the forms that result may, 
when compared, suggest the solution of the direct problem".  

For their study, they choose equations of the form 
\bea 
\sin(y)\,\sin(my)=a\sin(x)\,\sin(nx)+b\,,
\label{NP1}
\eea
where $m<1$ and $n<1$ are positive constants, mostly taken to be rational with denominators no greater than $11$,
while $a$ and $b$ are more general real
constants.

In a series of 24 plates,  they present the curves for over 100 different cases, each one obtained by first 
calculating the coordinates of a sufficiently large
number of points to determine a corresponding hand-drawn 
graph accurately.  This in turn  requires the determination of all solutions of a large number of 
equations of the form
\bea
\sin(x)\,\sin(mx)=  c\,
\label{NP2}
\eea
``by trial and error, a very simple process, but when often repeated quite tedious" \cite{newton_phillips}.  
They note that  as   the values of the constants in \eqref{NP1} are varied, the  curves can close or open to form a 
multitude of patterns, while always periodic in $x$ and $y$ for rational $m$ and $n$.  As  reviewer R.M. notes
when the paper is reported in {\em Nature} later in 1875
\cite{nature},
``These forms $\dots$ are all symmetrical, and much resemble carpet patterns.  
The tract is an interesting evidence of the 
patience and skill at draughtmanship of the authors".  

 For cases with $m=n=1/3$ and $a=-1$, 
they show in their Figs. 92--94
the graphs with  $b= 0.56$, $b=0$ and $b=-0.44$, and in  their Fig. 95,
the graphs of all three together.   Fig. 5 shows a copy of the part of the  plate from their paper containing these figures. 

\begin{figure}[htp]
\begin{center}
\includegraphics[width=\linewidth,clip, scale=1.0]{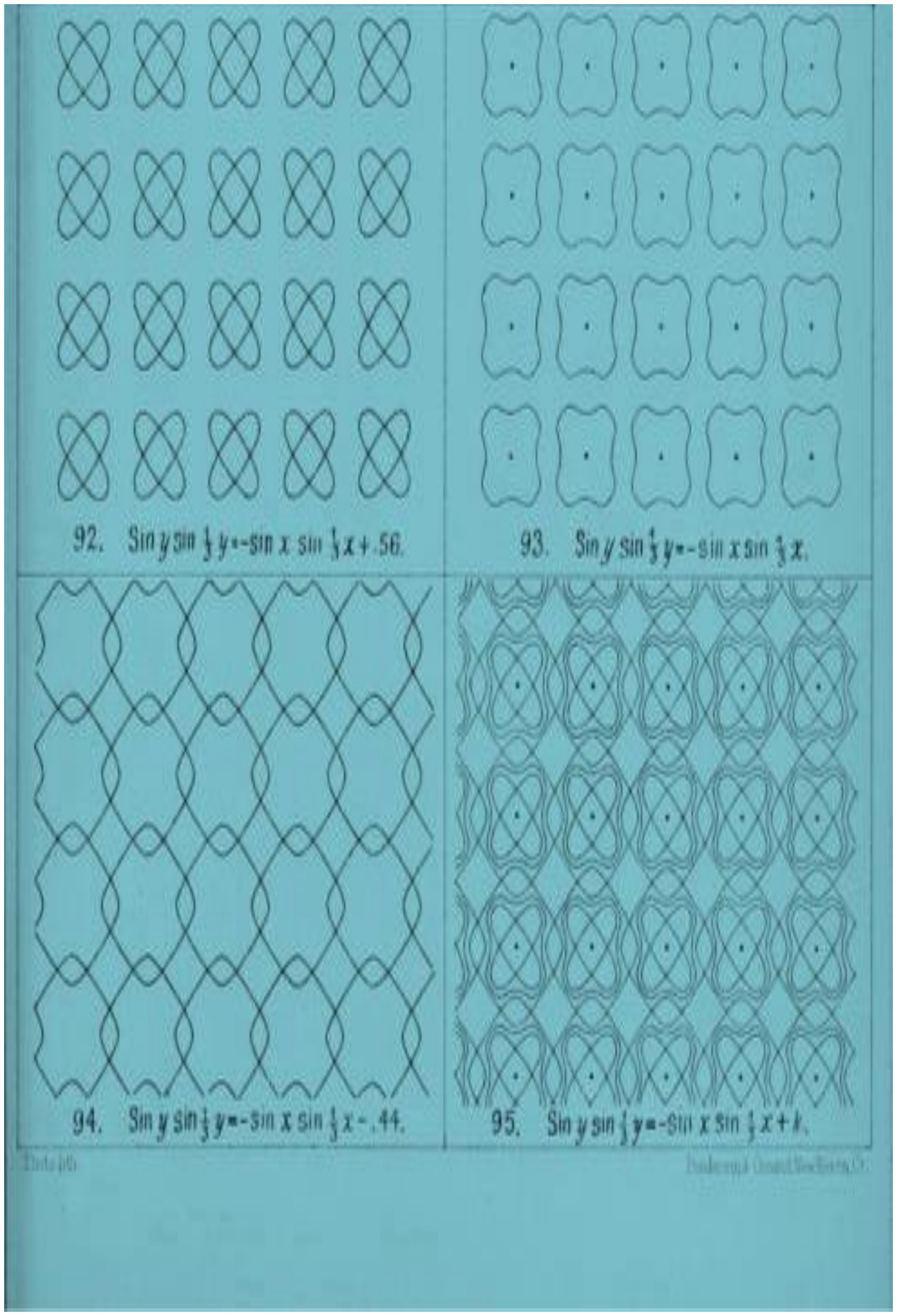}
\setlength{\abovecaptionskip}{-11pt}
\caption{Part of plate showing   Newton and Phillips' Figs. 92--95, courtesy of CAAS.
}

\end{center}
\end{figure}

Newton and Phillips
recognise that the third case is a limiting special case, as can be seen from their Fig. 95,
though they do not comment on that specifically in their paper.  
It is not clear whether they have shown, perhaps by a calculation like that  in Appendix A below, the critical 
nature of the value $b=-7/16$, 
to which $b=-0.44$ is an approximation.  
Perhaps they have calculated this value to two decimal places by some approximation scheme, 
or simply by plotting examples very accurately.  

Their Figs. 136--146 show the graphs in some cases with $m=1/2$ and $n=1/3$, in particular the cases with 
$a=1.37$ and with $b=0.60$, $b=0$ and $b=-0.77$, shown in  their Figs. 144--146 and  here in Fig.6.

 \begin{figure}[htp]
\centering
\includegraphics[width=\linewidth, clip, scale=1.0]{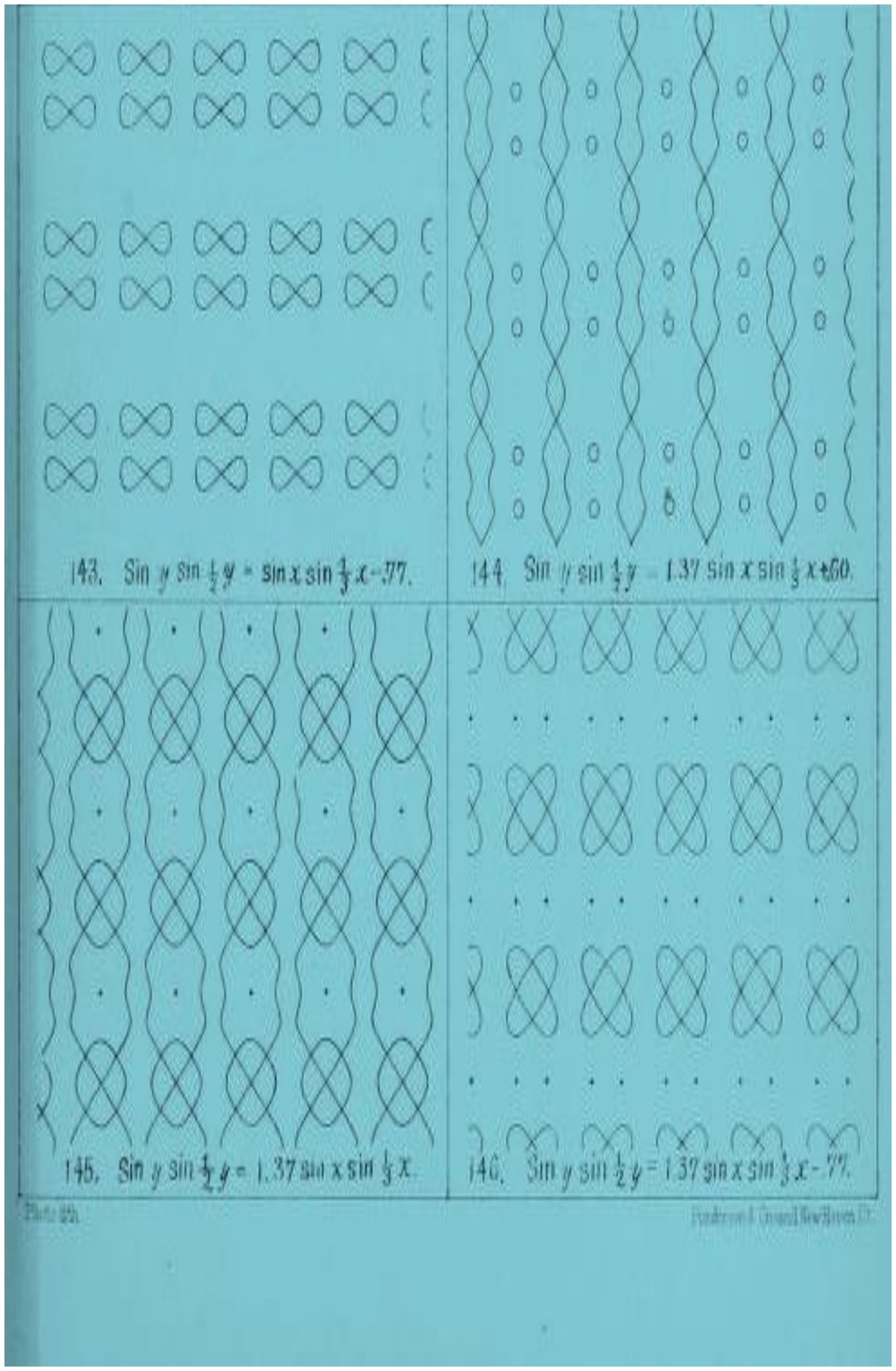}
\setlength{\abovecaptionskip}{-11pt}
\caption{Part of plate showing  Newton and Phillips' Figs. 144--146, courtesy of CAAS.
}
\end{figure}

Again, it is not known  if the authors have carried out as in Appendix B,
a calculation  for the case with $b=0$ to obtain the critical value
$a=64\sqrt{3}/81$, to which $a=1.37$ is an approximation, again accurate to two decimal places.  

The paper of Newton and Phillips will be noted in {\em Nature} \cite{nature} and  {\em Popular Science} 
\cite{popular_science}, and the
authors
will talk separately about aspects of their work at the Detroit meeting of  the AAAS later in 1875 \cite{AAAS_newton,AAAS_phillips}.
 
 Meanwhile, across the Atlantic, 35-year-old Friedrich Wilhelm Georg Kohlrausch has just arrived in 
 the German city of W\"urzburg  to
take up his position as Professor of Physics at JMU W\"urzburg.  There he will continue to build
an 
international reputation for research in electrochemistry and for the development of  laboratory techniques and equipment,
before moving on later to professorial positions in Strassburg   
in 1888 and then Berlin in 1900 \cite{goodwin}.  He too will be elected a Foreign Member of the Royal Society of London, in 1895.
His widely acclaimed textbook on the methods of experimental physics
has already appeared in 1870.  Translated into several languages and with many revisions, it
will continue to guide beginning experimental 
physicists well into the 21st Century, after reaching its 24th Edition \cite{leitfaden}. 

 Joining Kohlrausch in W\"urzburg in 1875 as Research Assistant and PhD student is a brilliant young Czech,  
 25-year-old Vincenc (\v{C}en\v{e}k) Strouhal from Prague, who has completed his undergraduate studies  at Charles University,
 with Ernst Mach as one of his teachers.  
In 1878, Strouhal
 will complete a ground-breaking thesis under Kohlrausch's guidance, 
 on the ``purring tones" that are emitted when thin stretched wires are subjected to cross-winds, 
 leading to 
 ``vortex shedding". 
His published results \cite{strouhal_paper} will be extensively
cited in years to come  (see for example the review \cite{sarpkaya}), and the dimensionless quantity that partly characterizes the phenomenon will come to be known as the ``Strouhal number".

But in 1875, Kohlrausch has been given the challenge of designing a new physics building on land recently acquired near 
what 
is now the R\"ontgen Ring in W\"urzburg, and young Strouhal is keen to participate in the project.  
In Prague he has been an enthusiastic student 
of mathematics, becoming treasurer of the Union of Czech Mathematicians, later to become the 
Union of Czech Mathematicians and Physicists, and he
has already presented  to the Union members, when aged only 21,  a series of eight talks \cite{novak}
on Gauss' famous study of curved surfaces \cite{gauss}.  

Strouhal obtains a copy of Newton and Phillips' paper,
possibly after seeing it  
listed in {\em Nature} \cite{nature}, where the reference to 
``carpet patterns" may have given him the idea to use some of the patterns in a tiled entrance to the new building.  
He convinces Kohlrausch
to make some money available for this purpose, and chooses the two patterns shown in Newton and Phillips'  Figs.
94 and 145, essentially the same as the computer-generated figures shown  in Figs. 3 and 4.
As already noted, the chosen patterns correspond to two particular formulas 
of the form \eqref{NP1}, the first with $m=n=1/3$, $a=-1$ and $b=-0.44$, and the second with $m=1/2$, $n=1/3$, 
$a=1.37$ and $b=0$.

Now Strouhal has to redraw each curve carefully, on a grid one period long, by 
one period wide, to define two tile-designs as clearly seen  in Figs. 1 and 2.   Thus each of the  tiles forming the inner pattern 
is  square, showing one period  in each direction  of the curves determined by the first formula,
with each of $x$ and $y$ running from $-3\pi/2$ to $3\pi/2$.   Similarly, each of the  tiles forming the outer, edging pattern
is  rectangular, showing one period  in each direction  of the curves determined by the second formula,
with $x$ running from  $-3\pi/2$ to $3\pi/2$ and $y$ running from   $-2\pi$ to $2\pi$.  The length scales for each
design are adjusted so the tiles will fit neatly together, as in Figs. 1 and 2. (The design for the corner tiles
is cleverly constructed from two parts of the second pattern.)

While redrawing the curves, Strouhal  notices that the critical values $b=-0.44$ for 
the first pattern and $a=1.37$ for the second as used by Newton and Phillips, are more accurately
given by $b=-0.4375$ and $a=1.3685$.   Perhaps he does the calculation of the exact values
$b=-7/16$, $a=64\sqrt{3}/81$ as in Appendices A and B below, 
but only records the values accurate to 4 decimal places.  
Or perhaps he uses an approximation scheme as Newton and Phillips may 
have done, and determines the values more accurately than they did, but still not exactly.  In truth the
changes in the curves resulting from use of the exact or either choice of approximate values are almost imperceptible.

In any event, Strouhal is sufficiently pleased with 
his efforts to have his improved
formulas fired onto tiles to accompany the two patterns, which he combines as shown in Figs. 1 and 2. 

Under the supervision of the engineer, Hrn. Ubach, the tiles are suitably colored and constructed
by the famous company of Villeroy and Boch in Mettlach, Germany,  and
then laid in the foyer of the new building, which is completed in 1879.

In 1882 Strouhal will return to Charles University in Prague as one of its founding professors of physics, 
and in his turn will have the opportunity to
contribute to the design of a new home for his discipline.  There he will  continue to build his own distinguished career.

\section{Concluding remarks}

I have no evidence that Newton or Phillips ever learned of the tiled floor in W\"urzburg,  
through contact with  Kohlrausch or Strouhal, or 
by some other means. 
But  I like
to imagine that well after their careers in New Haven and Prague were established,   
Andrew Phillips and Vincenc Strouhal met in W\"urzburg 
and admired the results of their separate endeavours
as PhD students on opposite  sides of the Atlantic.  

Phillips died in 1915 and Strouhal in 1922, but the tiled floor, complete with Newton and Phillips' formulas 
as amended by Strouhal, has survived to the present day.
This is not only a   
testament to the handiwork of the manufacturers Villeroy and Boch under the supervision of the engineer Hrn. Ubach, 
but as journalist Ernst N\"oth  rather grandly 
remarked 90 years after the floor was laid \cite{main_post_1971},
it also ``shows that mathematical truths are not affected by the passage of generations, governments, 
wars and fires, even when they are trodden underfoot by students, assistants and professors over many decades" (my translation).  

And I would add as a final remark: What a marvelous way to publish mathematics!

\renewcommand{\theequation}{A\arabic{equation}}
\setcounter{equation}{0} 
\section*{Appendix A:  The first formula}
It is convenient to rewrite the formulas defining the inner and outer patterns as

\bea 
(A) \quad F(y)+H(x)-b=0\quad  {\rm and}\quad   (B)\quad G(y)=a H(x)\,,
\label{formulas2}
\eea
where
\bea
F(y)=\sin(y)\,\sin(y/3)\,,\quad G(y)=\sin(y)\,\sin(y/2)\,,
\mea\mea
 H(x)=\sin(x)\,\sin(x/3)\,.\qquad\qquad\qquad
\label{formulas3}
\eea
The graphs of the periodic functions $F$, $G$ and $H$
are shown in Fig. 7.
\begin{figure}[htb]
\begin{center}
\includegraphics[width=\linewidth, height=4in, trim=0.5in 3.0in 0.1in 3.5in, clip, scale=0.5]{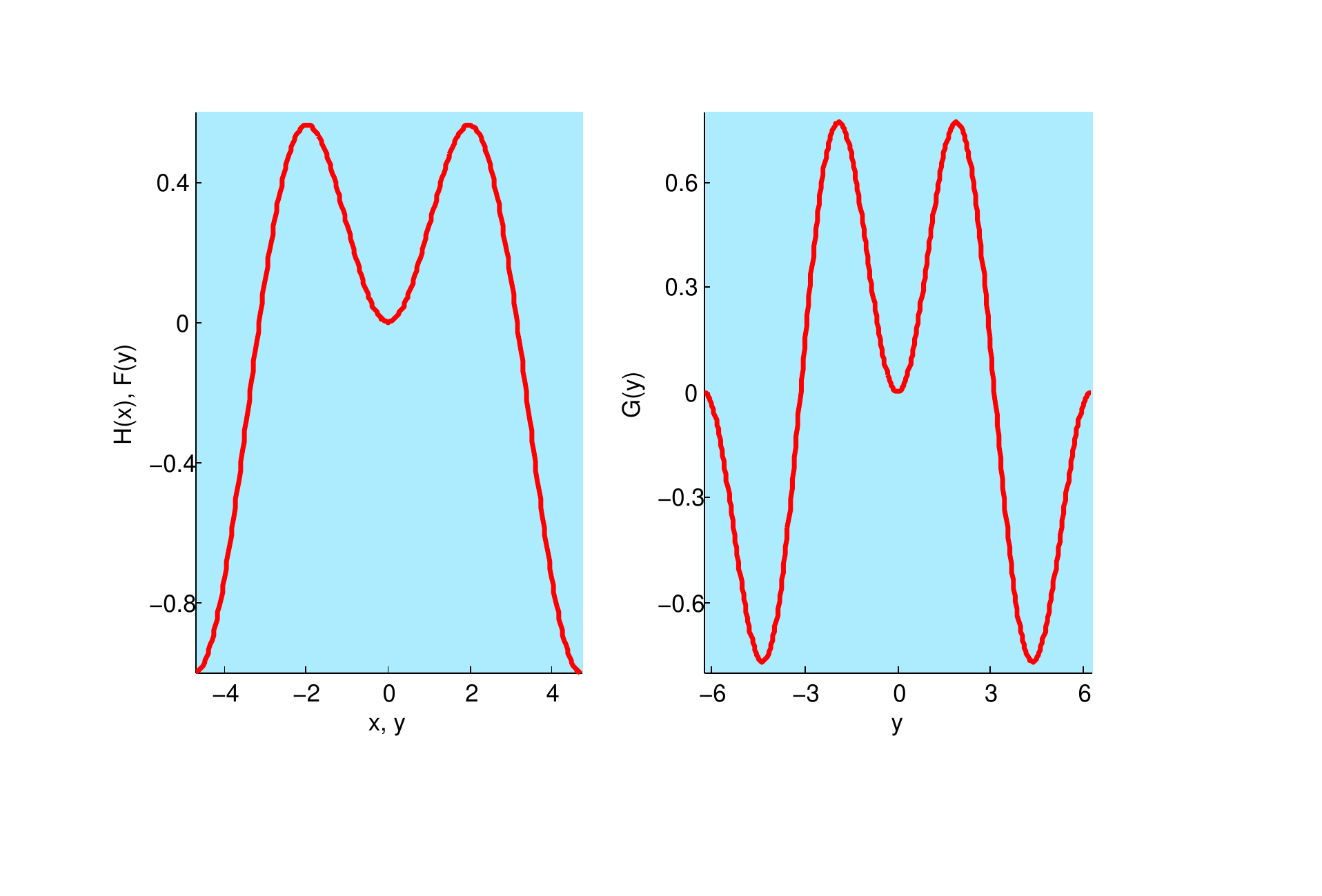}
\end{center}
\setlength{\abovecaptionskip}{-11pt}
\caption{On the left, graph of $H(x)$ and $F(y)$ over a period from $-3\pi/2$ to $3\pi/2$.
On the right, graph of G(y) over a period from $-2\pi$ to $2\pi$.}
\end{figure}

Newton and Phillips used $b=-0.44$ and $a=1.37$ to construct their Figs. 94 and 145, respectively, whereas Strouhal  put
$b=-0.4375$ and $a=1.3685$ in his formulas \eqref{formulas1}.  Here and in Appendix B 
we consider slightly more general values. 

If we consider a chain of  figures in the $XY-$plane  determined by formula $(A)$ in \eqref{formulas2},  
with $b$ values decreasing from
$b=-0.40$ to $b=-0.48$ for example, as shown from L to R in Fig. 8, we see in the middle figure that the curves close to form
an especially attractive pattern when 
$b=b_0\approx  -0.4375\approx  -0.44$, as chosen in \eqref{formulas1} and \cite{newton_phillips}. 
To determine this critical value exactly, we focus attention on $b>b_0$ and
in particular
on the two points where $dy/dx=0$, at $x=x_0\approx 2$ and $y\approx 5$.

\begin{figure}[htb]
\begin{center}
\includegraphics[width=\linewidth, height=4in, trim=0.5in 3.3in 0.5in 3.8in, clip, scale=0.7]{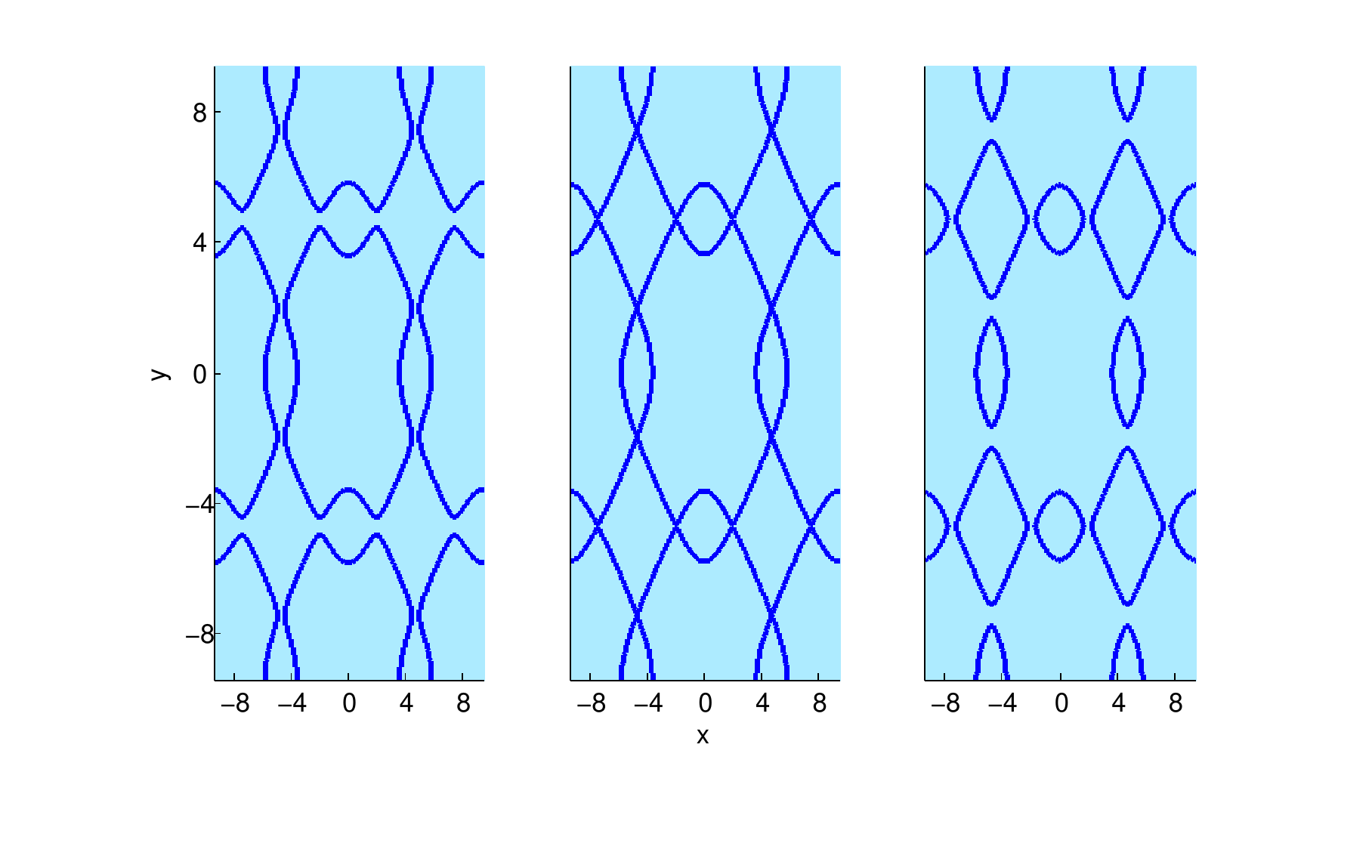}
\end{center}
\setlength{\abovecaptionskip}{-51pt}
\caption{Patterns determined by Formula $(A)$ with, from L to R,  $b=-0.40$, 
$b=-0.4375$, and $b=-0.48$.}
\end{figure}

 Firstly, we find $x_0$ from
 \bea
 F'(y)\,\frac{dy}{dx}=-H'(x)=\cos(x)\,\sin(x/3)+\third \sin(x)\,\cos(x/3)=0\,,
 \label{Acalc1}
 \eea
giving
\bea
3\tan(x/3)=-\tan(x)
\label{Acalc2}
\eea
and hence, with $t=\tan(x/3)$, 
\bea
3t=-\,(3t-t^3)/(1-3t^2)\,,\quad (\Rightarrow t=0, \sqrt{3/5} \quad\!\! {\rm or}\quad \!\!\! -\sqrt{3/5})\,.
\label{Acalc3}
\eea
The root of interest is $t=\sqrt{3/5}$, giving 
\bea
x_0=3\,\arctan\left (\sqrt{3/5}\right)\approx 1.9772\,.
\label{Acalc6}
\eea
Now we find
\bea
H(x_0)=\sin(x_0/3)\,\sin(x_0)=s_0(3s_0-4s_0^3)\,,\quad s_0=\sin(x_0/3)\,,
\label{Acalc7}
\eea
and use 
$t_0=s_0/\sqrt{1-s_0^2}$
to get
\bea
s_0^2=t_0^2/(1+t_0^2)=3/8\,,
\label{Acalc9}
\eea
and hence
\bea
H(x_0)=9/16\,.
\label{Acalc10}
\eea
From formula $(A)$ we then have at $x=x_0$
\bea
\sin(y)\,\sin(y/3)=-\,H(x_0)+b=\gamma,\,\, {\rm say},
\label{Acalc11}
\eea
or
\bea
S\,(3S-4S^3)=\gamma\,,\quad S=\sin(y/3)\,,
\label{Acalc12}
\eea
implying
\bea
S^2=(3\pm\sqrt{9-16\gamma})/8\,.
\label{Acalc14}
\eea
With $b<0$, we have $\gamma<0$ from \eqref{Acalc10} and \eqref{Acalc11}, and it follows 
that only the upper sign is relevant in \eqref{Acalc14}.
Because $S^2\leq 1$, it then follows using \eqref{Acalc10} that
\bea
3+\sqrt{9-16\gamma}\leq 8\quad \left(\Rightarrow-16\gamma\leq 16\,\,\Rightarrow b\geq -7/16\right)\,.
\label{Acalc16}
\eea
For each $b>-7/16$, it now follows that $S^2=\sin(y/3)^2<1$, leading to two possible values of $y$ 
with $dy/dx=0$ and with $0<y<3\pi$.
But when
$b=-7/16$, it follows that  $y=3\pi/2$ is the only possible value in this interval.  Thus the critical value  
$b_0$, when the two $y$ values with $dy/dx=0$ collapse to one, is the value given by Strouhal,
\bea
b_0=-7/16 =- 0.4375\,.
\label{Acalc17}
\eea

\renewcommand{\theequation}{B\arabic{equation}}
\setcounter{equation}{0} 
\section*{Appendix B:  The   second formula}
Now we consider a chain of figures determined by formula $(B)$  in \eqref{formulas2}, with $a$ values increasing
from $1.33$ to $1.41$ as shown in Fig. 9, and this time concentrate on the points near $x=2$, $y=2$ 
where $dy/dx=0$.

\begin{figure}[htb]
\begin{center}
\includegraphics[width=\linewidth, height=4in, trim=0.5in 3.3in 0.5in 3.8in, clip, scale=0.7]{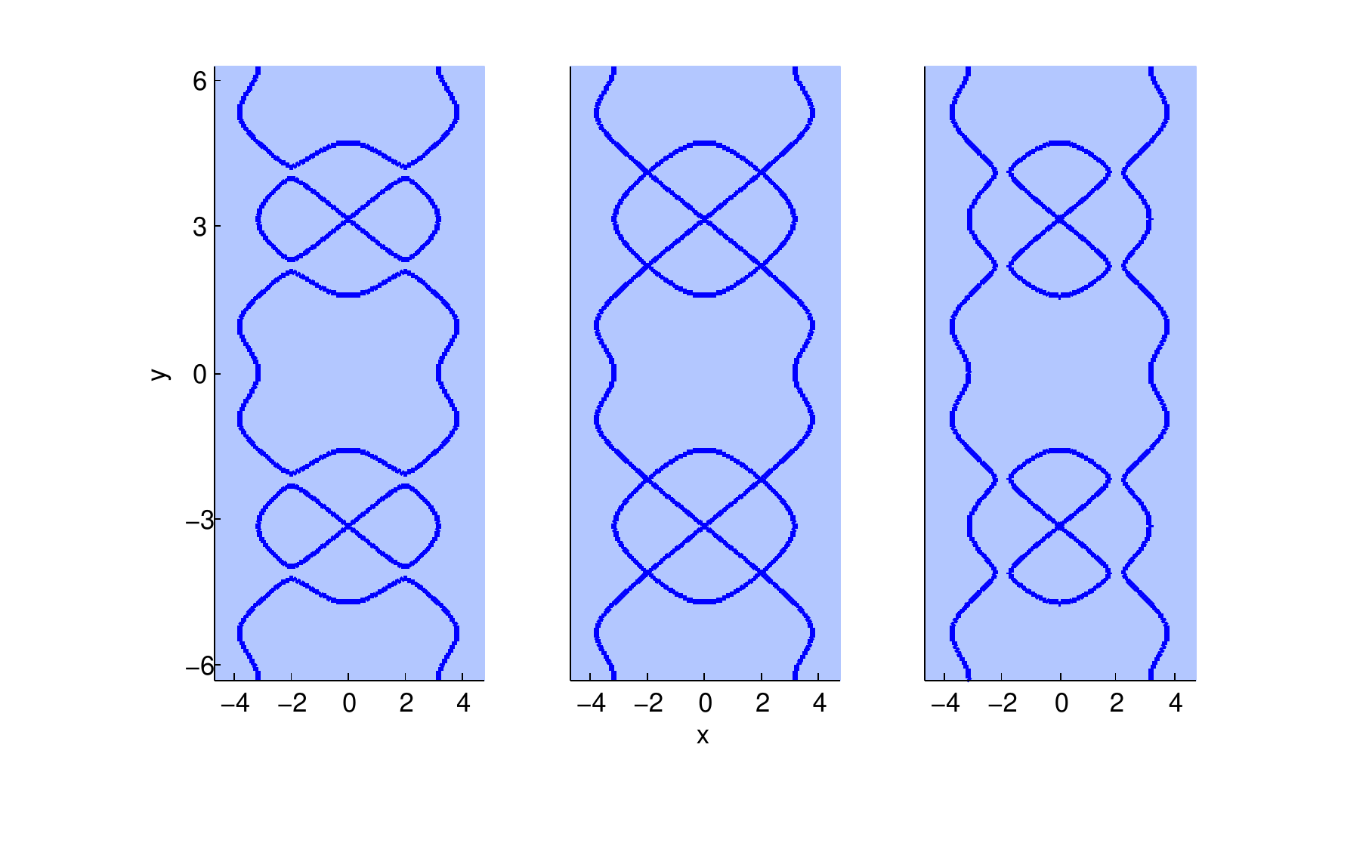}
\end{center}
\setlength{\abovecaptionskip}{-51pt}
\caption{Patterns determined by Formula $(B)$ with, from L to R,  $a=1.30$, 
$a=1.36853$, and $a=1.44$.}
\end{figure}

We see again a coincidence of these two points, this time when $a = a_0\approx 1.3685\approx 1.37$, 
the values chosen in \eqref{formulas1} and \cite{newton_phillips}.  
To determine the exact value of $a_0$, we note firstly that for general $a$, the $x$-cordinate of these points is 
again  $x=x_0$ as in \eqref{Acalc6}, so the $y$-coordinates of these points are determined by
\bea
G(y)=a\,H(x_0)=9a/16=\tau,\,\, {\rm say}.
\label{Bcalc1}
\eea
If we consider the graph of $G(y)$ as in Fig. 7, we see that there is a single maximum in the range of $y$-values of interest 
(say  $0<y<\pi$), occurring at a value $y=y_0$ to be determined.  Denoting  $G(y_0)$ by $ G_{{\rm max.}}$,
we see that 
there are two roots of \eqref{Bcalc1}  in the designated range provided $\tau <   G_{{\rm max.}}$, and just one root if 
$\tau =  G_{{\rm max.}}$.

 To calculate  $ G_{{\rm max.}}$,  we use
 \bea
 G'(y)=\cos(y)\,\sin(y/2)+\half\sin(y)\,\cos(y/2)=0 \,\, {\rm at} \,\, y=y_0\,,
 \label{Bcalc2}
 \eea 
giving
\bea
(1-2\sigma^2)\sigma+\sigma(1-\sigma^2) =0\,,\quad \sigma=\sin(y_0/2)\,,
\label{Bcalc3}
\eea
and hence (noting the allowed range of $y$-values) 
\bea
\sigma=\sqrt{2/3}\,.
\label{Bcalc4}
\eea
Now $\cos(y_0/2)=\sqrt{1-\sigma^2}=1/\sqrt{3}$, and hence
\bea
G_{{\rm max.}}=G(y_0)=2\sin^2(y_0/2)\,\cos(y_0/2)=4/3\sqrt{3}\approx 0.770\,.
\label{Bcalc5}
\eea

Because the critical value of $\tau$ is given by $G_{{\rm max.}}$, it now follows from \eqref{Bcalc1} that the critical
value of $a$ is given by
\bea
a_0= 64\sqrt{3}/81=1.36853...
\label{Bcalc6}
\eea

\ni
{\bf Acknowledgements:} 
My thanks to Vincent Hart, Ludvik Bass, Peter Jarvis, Gunnar Bartsch, Anja Schl\"omerkemper,
Wolfgang Schlegel, Steve Webb,  Robert Grebner and  Matthias Reichling for their various inputs and encouragement.


\begin{thebibliography}{99}
\bibitem{roentgen}
Riesz, P.B., The life of Wilhelm Conrad Roentgen, {\em Am. J. Roentgenol.} {\bf 165} (1995), 1533--1537.  

\bibitem{bracken1}
Bracken, A.J., The mystery of the strange formulae,  {\em Phys. World}, (Oct. 2016, p. 22). 

\bibitem{schlegel}
Schlegel, W., The tale of the tiles, {\em Phys. World}, (Dec. 2016, p. 21).

\bibitem{main_post_1971}
N\"oth, E., {\em sin x und sin y unter den F\"ussen}, (Main-Post, W\"urzburg, July 2, 1971). 


\bibitem{matlab}
{\em MATLAB} (MathWorks, Natick MA, 2016).

\bibitem{bartsch}
Bartsch, G., {\em  R\"atselhafte Spuren in R\"ontgens Labor} (einBLICK, Presse-und \"Offentlichkartsarbeit, JMU W\"urzburg),
Dec. 13, 2016. 

http://www.presse.uni-wuerzburg.de/aktuell/einblick/einblick\_archiv
/ausgaben\_ab\_2013/
liste/page/2/zeitraum/2016/12/?tx\_news\_pi1
\%5B controller\%5D=News\&cHash=59ea6b3f02997ba9c44e1fdee0f371a4

\bibitem{main_post2}

{\em Das Fussboden-Rätsel in Röntgens Labor}
(Main-Post, W\"urzburg, Dec. 19, 2016).

http://www.mainpost.de/regional/wuerzburg/Allgemeine-nicht-fachgebundene-Universitaeten-Mathematik-Mathematiker-Physik-Roentgen-Roentgen-Gedaechtnisstaette;art735,9448939


\bibitem{newton_phillips}
Newton, H.A. and Phillips, A.W., On the Transcendental Curves $\sin y \sin my =a \sin x \sin nx +b$, {\em Trans. Conn. Acad. Arts Sci.} {\bf 3} (1874--1878), 97--107 (with 24 plates).

http://www.biodiversitylibrary.org/item/88413\#page/117/mode/1up


\bibitem{phillips}
Phillips, Andrew W., Biography: Hubert Anson Newton, {\em  Amer. Math. Monthly} {\bf  4} (no. 3)  (1897),  67--71.

\bibitem{gibbs}
Gibbs, J. Willard, Memoir of Hubert Anson Newton, 1830--1896,  {Nat. Acad. Sc. USA}.

\bibitem{wright}
Wright, H.P., Early ideals and their realization,  
in {\em Andrew Wheeler Phillips} (Tuttle, Morehouse \& Taylor, New Haven, 1915), pp. 5--13.

https://babel.hathitrust.org/cgi/pt?id=hvd.hn2k3q;view=1up;seq=13

\bibitem{roberts}
Roberts, S.,  On three-bar motion in plane space, {\em Proc. Lond. Math. Soc.} {\bf s1--7} (1875), 15--23.

\bibitem{cayley} 
Cayley, A., On three-bar motion, {\em Proc. Lond. Math. Soc.} {\bf s1--7}  (1876), 136--166.


\bibitem{phillips5}
Phillips, A.W. and Beebe, W., {\em Graphic Algebra, or Geometrical Interpretation of the 
Theory of Equations of One Unknown Quantity}, (H. Holt, NewYork, 1904).  




\bibitem{nature}
Our Book Shelf, {\em Nature} {\bf 13} (no. 338) (1876), 483. 

 \bibitem{popular_science}
 Miscellany, {\em Popular Science Monthly} {\bf 8} (1875), 121.

\bibitem{AAAS_newton}
Newton, H.A., Algebraic curves expressed in trigonometric equations,  
AAAS, 24th Meeting, Detroit, Aug. 11, 1875, {\em Amer. Chemist} (Sept. 1875), 103.

https://books.google.com.au/books?id=ZSSduuWOpkAC\&pg=PA103

\bibitem{AAAS_phillips}
Phillips, A.W., On certain transcendental curves, AAAS, 24th Meeting, Detroit, Aug. 11, 1875, 
{\em Amer. Chemist} (Sept. 1875), 103.

https://books.google.com.au/books?id=ZSSduuWOpkAC\&pg=PA103


\bibitem{goodwin}
Goodwin, H.M. (Transl. and Ed.), Biographical sketch, in {\em The Fundamental Laws of Electrolytic Conduction}
(Harper \& Bros., NY, 1899), pp. 92-3.
  
http://www.archive.org/stream

/fundamentallawso00goodrich\#page/92/mode/2up

\bibitem{leitfaden}
Kohlrausch, F.W.G., {\em Leitfaden der Praktische Physik} (B.G. Teubner, Leipzig, 1870);
{\em Praktische Physik}, 24th Ed. (Springer Vieweg, Berlin, 1996).


\bibitem{strouhal_paper}
Strouhal, V., \"Uber eine besondere Art der Tonerregung, {\em Ann. d. Phys.} {\bf 241 (10)} (1878), 216--251.

\bibitem{sarpkaya}
Sarpkaya, T., Vortex-induced oscillations: a selective review, {\em J. Appl. Mech.} {\bf 46} (1979), 241--258.   

\bibitem{novak}
Nov\'ak, V., \v{C}en\v{e}k Strouhal, {\em \v{C}asopis pro p\v{e}stov\'an\'i mathematiky a fysiky 
(J. for the Promotion of Mathematics and Physics)}  {\bf 39} (1910), 369--383.  

\bibitem{gauss}
Gauss, C.F., {\em  Disquisitiones generales circa superficies curvas} (Typis Dieterichianis, G\"ottingen, 1828). 


\end{thebibliography}
\end{document}